\documentclass[conference]{IEEEtran}
\usepackage{cite}
\usepackage{amsmath,amssymb,amsfonts}
\usepackage{algorithmic}
\usepackage{graphicx}
\usepackage{textcomp}
\usepackage{xcolor}
\usepackage{fancyhdr}
\usepackage[hyphens]{url}

\usepackage{caption}
\usepackage{subcaption}

\newcommand{\transmap}{TransforMAP}
\def\BibTeX{{\rm B\kern-.05em{\sc i\kern-.025em b}\kern-.08em
    T\kern-.1667em\lower.7ex\hbox{E}\kern-.125emX}}

\title{TransforMAP: Transformer for Memory Access Prediction}

\author{\IEEEauthorblockN{Pengmiao Zhang}
\IEEEauthorblockA{\textit{University of Southern California} \\
Los Angeles, USA \\
pengmiao@usc.edu}

\\
\IEEEauthorblockN{Anant V. Nori}
\IEEEauthorblockA{\textit{Processor Architecture Research Lab, Intel Labs} \\
Bangalore, India \\
anant.v.nori@intel.com}
\and
\and
\IEEEauthorblockN{Ajitesh Srivastava}
\IEEEauthorblockA{\textit{University of Southern California} \\
Los Angeles, USA \\
ajiteshs@usc.edu}
\\

\and
\IEEEauthorblockN{Rajgopal Kannan}
\IEEEauthorblockA{\textit{US Army Research Lab} \\
Los Angeles, USA \\
rajgopal.kannan.civ@mail.mil}
\\
\IEEEauthorblockN{Viktor K. Prasanna}
\IEEEauthorblockA{\textit{University of Southern California} \\
Los Angeles, USA \\
prasanna@usc.edu}

}

\begin{document}
\maketitle
\pagestyle{plain}


\begin{abstract}
Data Prefetching is a technique that can hide memory latency by fetching data before it is needed by a program. Prefetching relies on accurate memory access prediction, to which task machine learning based methods are increasingly applied. Unlike previous approaches that learn from deltas or offsets and perform one access prediction, we develop~\transmap, based on the powerful Transformer model, that can learn from the whole address space and perform multiple cache line predictions.
We propose to use the binary of memory addresses as model input, which avoids information loss and saves a token table in hardware. We design a block index bitmap to collect unordered future page offsets under the current page address as learning labels. As a result, our model can learn temporal patterns as well as spatial patterns within a page. In a practical implementation, this approach has the potential to hide prediction latency because it prefetches multiple cache lines likely to be used in a long horizon.
We show that our approach achieves 35.67\% MPKI improvement and 20.55\% IPC improvement in simulation, higher than state-of-the-art Best-Offset prefetcher and ISB prefetcher. 
\end{abstract}
\section{Introduction}
Memory access prediction is a very important task for data prefetching, which is a widely used technique for hiding memory latency and improving instructions per cycle (IPC). A prefetching process is a form of speculation that aims to predict the future data addresses and fetch the data before it is needed. Hardware prefetchers usually exploit obvious memory access patterns, such as the adjacent spatial locality and constant stride. For example, spatial Memory Streaming (SMS)~\cite{somogyi2006spatial} prefetcher identifies code-correlated spatial patterns and streams at run time and predicts future accesses using these patterns. Best-Offset prefetching approach proposed in~\cite{michaud2016best} predict offsets while taking into account of prefetching timeliness. Irregular Stream Buffer (ISB)~\cite{jain2013linearizing} learns temporally correlated memory accesses based on PC-localized stream. 

However, hardware prefetchers are incapable of learning latent trace patterns or providing generalized inference. Machine learning has provided insights into data prefetching. In \cite{Rahman2015-xk}, the authors propose to use logistic regression and decision tree models to maximize the effectiveness of existing hardware prefetchers in a system.~\cite{mempatterns} presents an extensive evaluation of recurrent neural networks in learning memory access patterns and demonstrates high performance in precision and recall. Some other works~\cite{peled2018neural, zeng2017long, braun2019understanding} also demonstrate the effectiveness of LSTM in memory access prediction.~\cite{srivastava2019predicting, srivastava2020memmap,zhang2020raop} use compact LSTM and meta-model techniques to reduce the model size pursuing to build a practical prefetcher. In~\cite{deepcache}, the authors first propose using Seq2seq modeling, based on LSTM Encoder-Decoder structure, to predict the future characteristics of an object for content caching and significantly boosts the number of cache hits. 


Due to the underlying grammar similarity between memory accesses and natural language, natural language processing (NLP) models are naturally applicable to learning accesses~\cite{srivastava2020memmap,zhang2021c}. The Transformer~\cite{vaswani2017attention}, a sequence model based on multi-head self-attention initially proposed for machine translation, has achieved huge success for sequence modeling tasks in many fields compared to traditional recurrent models. This suggests that self-attention might also well-suited to modeling memory access patterns. Our target is to bring the paradigm of memory access prediction for data prefetching under the Transformer architecture. Unlike most NLP problems that usually have clear labels and reasonable corpus size for learning, the data prefetching task is presenting two challenges we need to tackle: unfixed labeling, class explosion~\cite{shi2021hierarchical}, and latency of prediction. Unfixed labeling indicates that there is no ground truth that a prefetcher should prefetch a certain memory address because any address following the current access could be the labels. Class explosion indicates that the class space would be the same as the address space if the model input/output uses absolute memory address and the problem is formulated as classification. A 64-bit address space requires a model to predict one class out of tens of millions of classes. Latency of prediction indicates that the inference latency of ML-based prefetchers can be larger than traditional rule-based prefetchers, which can cause late prefetches and make the prefetching useless even for accurate predictions.

In this paper, we propose~\transmap, a Transformer-based memory access prediction framework, to tackle the above challenges. We use the same Encoder-Decoder structure as the Transformer model. On the encoder side, we propose to use the sequence of addresses in binary as input. This input format presents three advantages. First, comparing to deltas, the binarized address only incorporates a vocabulary size at 2, which largely reduces the input vocabulary space. Second, the vocabulary is fixed and needs no tokenization since class itself can be used as values in the computation, which avoids the word-to-index table in hardware implementation. Third, there is no information loss compared to using only deltas or offsets as inputs. On the decoder side, we propose a bitmap labeling technique to set the training label as unordered future $k$ page offsets under the same page as the current address. Since we only consider the pattern within a page, the vocabulary for the output is the offset space, e.g., 64 for a 12-bit page with a 6-bit block.~\transmap~can deal with an unfixed number of multiple block predictions and bitmap labeling enhances the model's capacity in predicting longer patterns, thus, offsets the computation latency.

\begin{figure*}
    \centering
    \includegraphics[width=\linewidth]{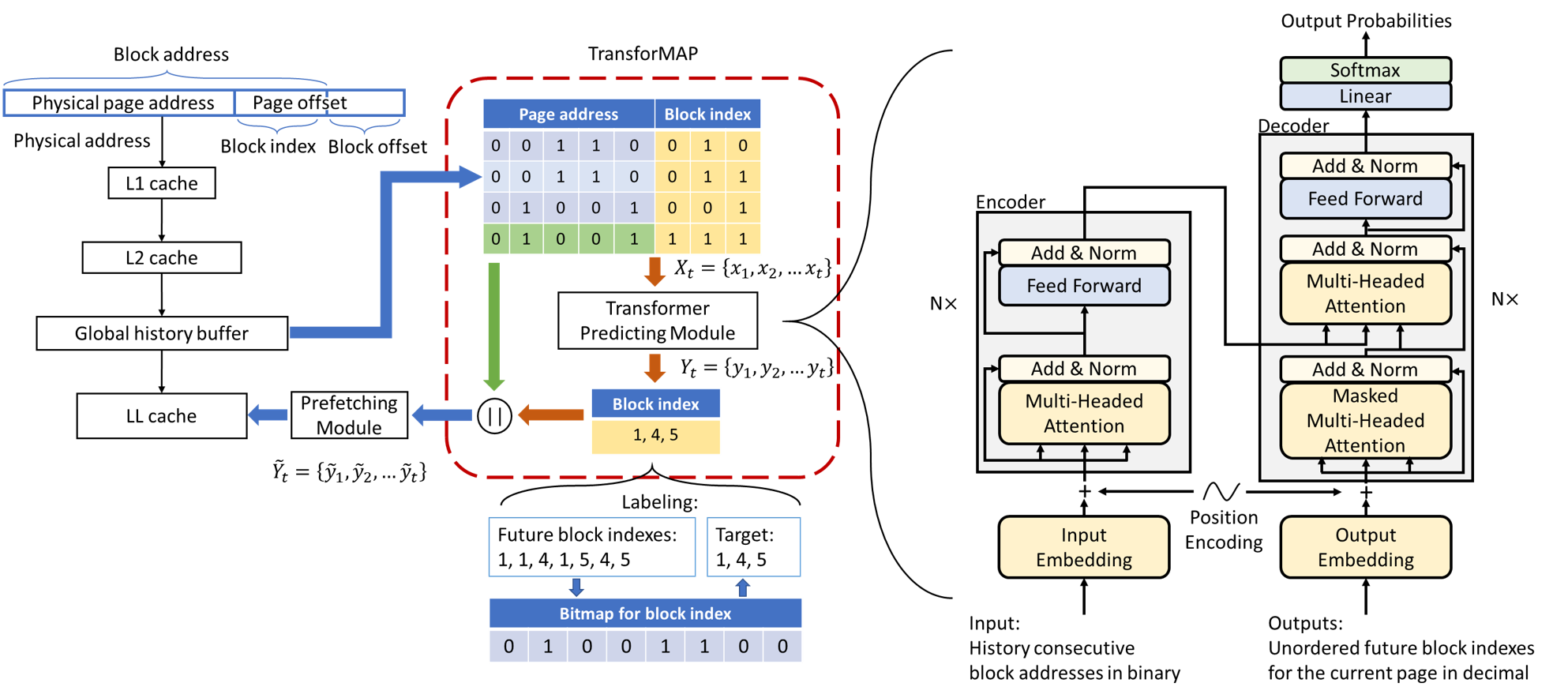}
    \caption{Overall architecture of~\transmap. We have an input sequence of history block addresses in binary $X_t = \{x_1,x_2, ..., x_t\}$ and output sequence the desired block index $Y_t=\{y_1, y_2, ..., y_k\}$ under the current page. The final address $\widetilde{Y_t}$ prediction is the concatenation of current page address, the predicted block index, and the block offset.}
    \label{fig:overall}
\end{figure*}

Our contribution can be summarized as follows:
\begin{itemize}
    \item We propose a Transformer-based framework for the task of multiple memory access predictions. 
    \item We propose to use the binary of addresses as model inputs, which solves the problem of class explosion without information loss compared to using deltas or offsets as inputs. It also avoids extra token tables in hardware implementation.
    \item We propose a bitmap labeling approach that collects unordered future page offsets under the same page address that avoids unnecessary repetitive prediction. It facilitates long-horizon prediction that offsets the misses caused by calculation latency from a practicality perspective.
    \item We evaluate our method using ChampSim simulator. Results show that~\transmap~achieves 35.67\% MPKI improvement and 20.55\% IPC improvement, outperforms Best-Offset prefetcher and ISB prefetcher.
     
\end{itemize}

\section{Related Work}

Several prior works have explored the application of machine learning algorithms on data prefetching and memory access prediction.~\cite{srivastava2019predicting} proposed a compact LSTM based prediction model and extensively studied the memory patterns, LSTM prediction performance, and online learning strategies. The authors train LSTM models with virtual memory access deltas and have achieved high accuracy. ~\cite{zhang2020raop} and~\cite{srivastava2020memmap} also leverages LSTM-based models and deals with virtual addresses, focusing more on the practicality of model implementation on hardware. However, consecutive virtual memory accesses patterns are more notable than the translated physical addresses, especially for the last-level caches where the memory accesses have been filtered by lower-level caches. Simple LSTM models may not be capable to learn the physical pattern in LLC so we resort to a more powerful model to tackle this challenge. 

~\cite{shi2021hierarchical} proposes the Voyager model for memory access prediction. This model predicts both page sequence and page offsets. The Voyager model uses two LSTM models to strengthen the model learning capacity. The authors use a simplified dot-product attention mechanism without scale factor to build a connection between input page embedding and input offset embedding. In this work, we will explore how a model with only attention mechanisms, without recurrent network structure, performs on the memory access prediction task.

\section{Approach}

\subsection{Overview of~\transmap}

Figure~\ref{fig:overall} illustrates the overall architecture of the proposed~\transmap~and how the model is applied in a hardware system. We try to leverage a state-of-the-art machine learning algorithm, the Transformer, to improve the cache hit. We treat the prefetching problem as sequence prediction and perform classification on page offsets. Because a prefetch must be in the unit of a block (or cache line), we can increase the granularity and consider only the block index space, the configuration is shown in the left top of figure~\ref{fig:overall}. 

\noindent\textbf{Problem Formulation} In abstract, let $ A_t = \{a_1, a_2, ..., a_t\}$ be the sequence of history block addresses at time $t$. Let $X_t = \{x_1,x_2, ..., x_t\}$ be the binary representation of $ A_t$, where $x_t=\{b_t^1, b_t^2,..., b_t^m,..., b_t^{m+n}\}$ represents the $m$-bit binary values for page address and $(m+n)$-bit binary values for block address at time $t$. Let $Y_t=\{y_1, y_2, ..., y_k\}$ be the sequence of $k$ outputs associated with the unordered future $k$ block index for the same page address. Our goal is to construct meaningful $X_t$ to $Y_t$ that are helpful in data prefetching. The final address prediction $\widetilde{Y_t}=\{\widetilde{y_1}, \widetilde{y_2}, ..., \widetilde{y_k}\}$ is the concatenation of current page address, the predicted block index, and the block offset. 

\subsection{Input Sequence}
Memory data fetching is on the unit of cache lines or blocks. Therefore, we only consider the address bits upper than the block offset, referred to as block addresses. To deal with the extremely large vocabulary of address space, $2^{64}$ for a 64-bit address, we use a binary vector to represent the absolute address value. There are three significant advantages of using the binary vector as input. First, the vocabulary can be largely reduced. For a $l$-bit physical address, while this method increases the input sequence length to $l\times$, it results in a $2^{l-1}\times$ reduction of vocabulary. Second, the vocabulary is fixed as two: 0 and 1, and they are numerical values. There is no need for tokenization and the input can be directly used for calculation. This is significant for hardware implementation because it saves an extra table storing the token dictionary and avoids the process of word-to-index conversion.
Third, there is no information loss compared to using only deltas or offsets as inputs. Binary vector is equivalent to the absolute address while deltas leverage the difference between memory accesses and offsets only consider part of the address. An inference from this advantage is that the model can learn from the whole address space and can handle an address that has never appeared before. The advantage is highly significant in memory access prediction task because of orders of magnitudes larger vocabulary size compared to natural language tasks.

\subsection{Bitmap Labeling}

We aim to predict unordered future block indexes under the same page as the current memory address. The label is collected from offline memory access traces. This labeling method is based on the hypothesis that memory access pattern within a page is more significant and easier to track. 

As is shown in the bottom part of figure~\ref{fig:overall}, we use a bitmap at the length of block index space to store our labels. For example, given a 3-bit block index, a bitmap at length 8 is required. The index of a bitmap is equivalent to the block index in the memory address. While the future block indexes appear in order, we only record the appearance of the block indexes in bitmap and ignore the order. The bitmap indexes with value 1 are set as model labels.
From an algorithm perspective, this method avoids repetitive prediction and decreases the complexity of output space. From a hardware perspective, this method facilitates the model to predict longer future accesses and offsets the near future miss caused by computation latency.

\subsection{Transformer Predicting Module}
We use the state-of-the-art machine learning algorithm in sequence modeling, the Transformer~\cite{vaswani2017attention}, to learn the mapping between history address sequences in binary and the future block indexes in decimal.

\subsubsection{Transformer Layers}
The right part of figure~\ref{fig:overall} shows the architecture of the Transformer model we use. We employ the model using an Encoder-Decoder structure similar to the original architecture. 

\noindent\textbf{Self-attention.} The scaled dot-product attention is defined as follows:

\begin{equation}
\operatorname{Attention}(Q, K, V)=\operatorname{softmax}\left(\frac{Q K^{T}}{\sqrt{d_{k}}}\right) V
\end{equation}

where $Q$ represents the queries, $K$ the keys, and $V$ the values. The self-attention operations take the embedding of items as input, and convert them to three matrices through linear projection, and feeds them into an attention layer. $d$ is the dimension of the layer input.

\noindent\textbf{Multi-headed attention.} One self-attention operation can be considered as one "head", we can apply multi-head attention operation as follows:

\begin{equation}
\begin{aligned}
\operatorname{MultiHead}(Q, K, V) &=\operatorname{Concat}\left(\operatorname{head}_{1}, \ldots, \text {head}_{\mathrm{h}}\right) W^{O} \\
\text { where head }_{\mathrm{i}} &=\text { Attention }\left(Q W_{i}^{Q}, K W_{i}^{K}, V W_{i}^{V}\right)
\end{aligned}
\end{equation}

where the projection matrics $W_{i}^{Q}, W_{i}^{K}, W_{i}^{V} \in \mathbb{R}^{d \times d}$ and h is the number of heads.

\noindent\textbf{Point-wise feed-forward.} Point-wise feed-forward network (FFN) is defined as follows:
\begin{equation}
\operatorname{FFN}(x)=\max \left(0, x W_{1}+b_{1}\right) W_{2}+b_{2}
\end{equation}

\noindent\textbf{Position encoding}
Because there are no recurrent steps in the self-attention layer, positional encodings are leveraged before both encoder and decoder to inject the orders of elements in a sequence to the model~\cite{gehring2017convolutional}. We use sine and cosine function for positional encoding:
\begin{equation}
\begin{aligned}
P E_{(p o s, 2 i)} &=\sin \left(p o s / 10000^{2 i / d_{\text {model }}}\right) \\
P E_{(p o s, 2 i+1)} &=\cos \left(p o s / 10000^{2 i / d_{\text {model }}}\right)
\end{aligned}
\end{equation}

\subsubsection{Loss Function} For our multi-class classification problem, we use cross-entropy loss defined as below:
\begin{equation}
L_{\log }(Y, P)=-\frac{1}{N} \sum_{i=0}^{N-1} \sum_{k=0}^{K-1} y_{i, k} \log p_{i, k}
\end{equation}
where $Y$ is the matrix of true labels, $P$ is the matrix of prediction probability, $y_{i,k}$ is the true value for the $i$th sample at class $k$, $p_{i,k}$ is the probability for the $i$th sample to be class $k$.

\subsubsection{Training}
We use the Adam optimizer~\cite{kingman2015adam} with $\beta_1 = 0.9$, $\beta_2 = 0.98$ and $\epsilon=10^{-9}$. Custom learning rate over the course of training is used as follows with $warmup\_steps=2000$:
\begin{equation}
\begin{aligned}
lrate=d_{\text {model}}^{-0.5}\cdot\min({step\_num}^{-0.5}, \\
step\_num \cdot warmup\_steps)
\end{aligned}
\end{equation}

\subsubsection{Inference}
We use beam search with beam size = 2 to find an output with a maximum likelihood. At time step 1, we feed the output embedding with a $<beginning>$ token and conduct beam search until the appearance of $<ending>$ token or the inference achieves the maximum length of output sequence. 

\subsection{Output Concatenation}
The model output is only the predicted future block indexes, we need to convert the prediction back to the absolute address that can be used by a prefetching module. Let the current address be $a_t$, one predicted block index $y_k$, the final predicted address for prefetching is:
\begin{equation}
\begin{aligned}
\widetilde{y_k}= (a_t \gg \log_2{(page\_size)} \ll block\_index + y_k)\\ \ll \log_2{(block\_size)} \\
block\_index = \log_2{(page\_size)}-\log_2{(block\_size)}
\end{aligned}
\end{equation}

\begin{figure}
\begin{subfigure}{.5\textwidth}
  \centering
  \includegraphics[width=\linewidth]{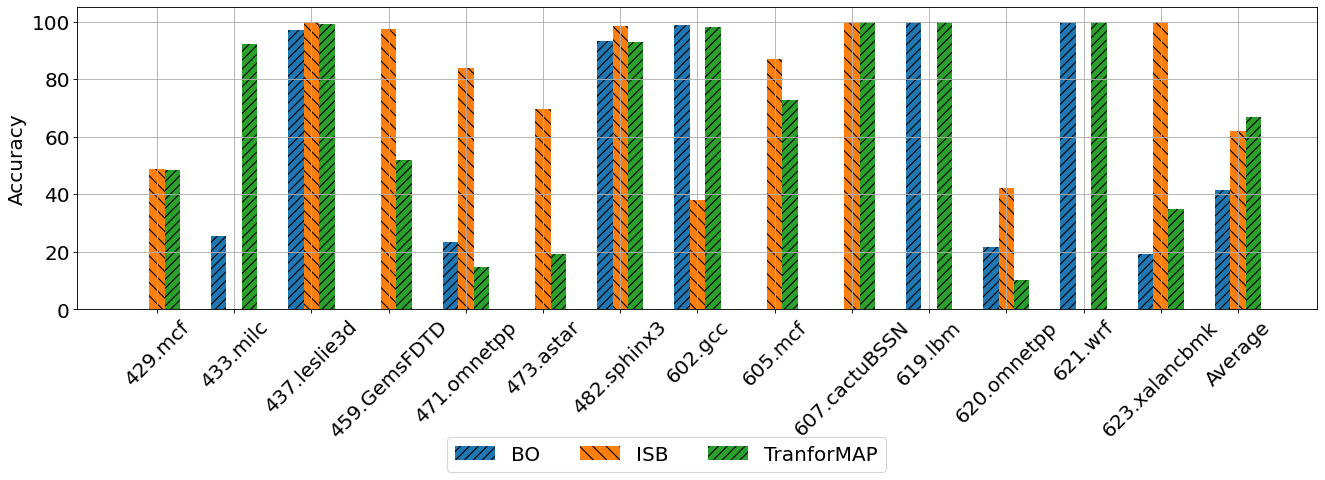}  
  \caption{Accuracy}
  \label{fig:sub-first}
\end{subfigure}
\begin{subfigure}{.5\textwidth}
  \centering
  \includegraphics[width=\linewidth]{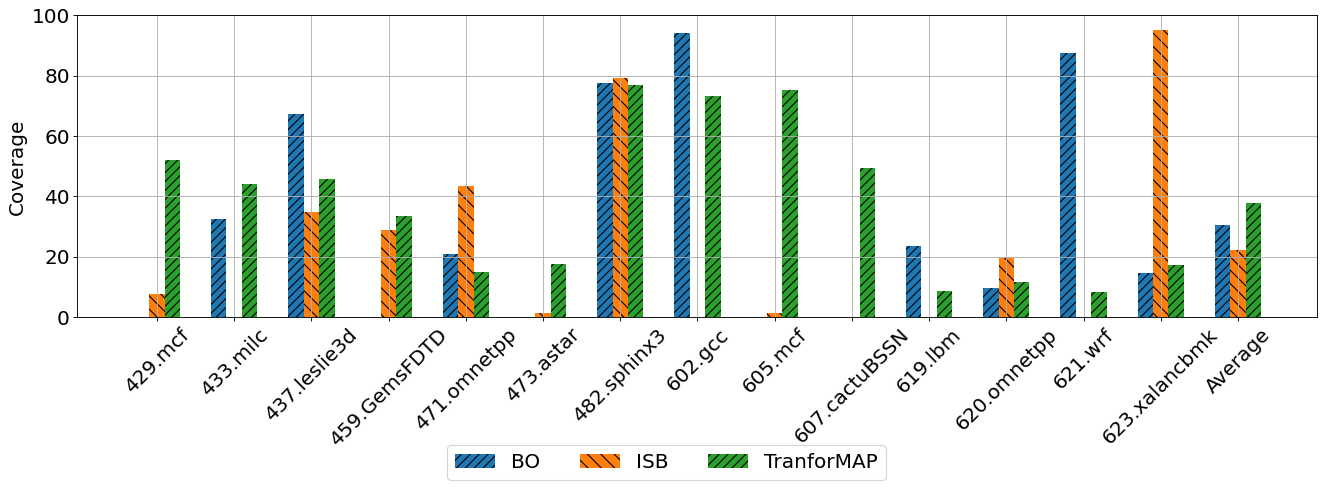}  
  \caption{Coverage}
  \label{fig:sub-second}
\end{subfigure}
\newline
\begin{subfigure}{.5\textwidth}
  \centering
  \includegraphics[width=\linewidth]{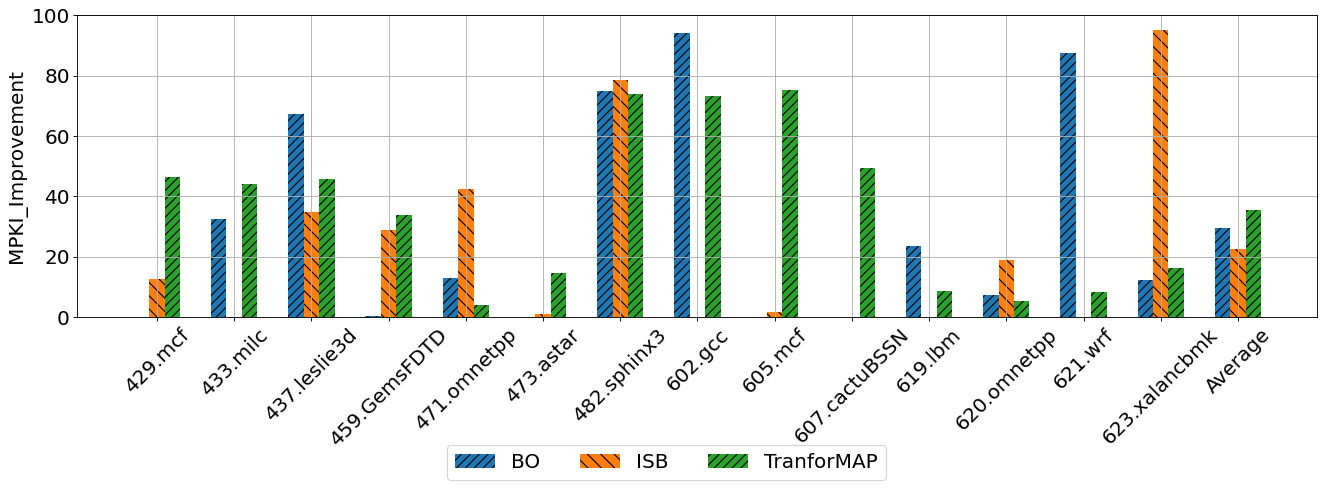}  
  \caption{MPKI improvement}
  \label{fig:sub-third}
\end{subfigure}
\begin{subfigure}{.5\textwidth}
  \centering
  \includegraphics[width=\linewidth]{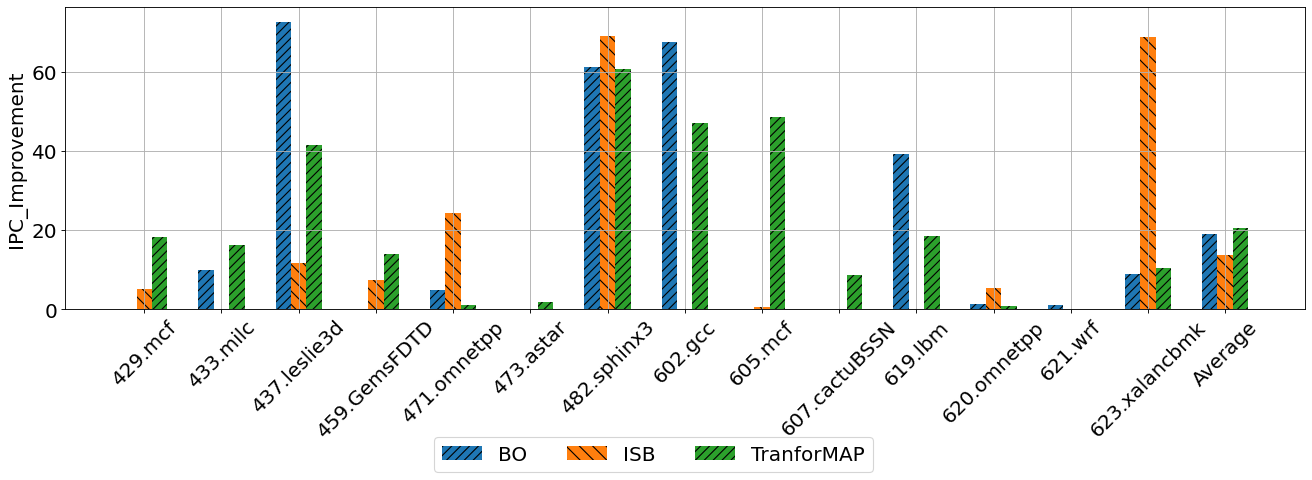}  
  \caption{IPC improvement}
  \label{fig:sub-fourth}
\end{subfigure}
\caption{Simulation results}
\label{fig:res}
\end{figure}

\section{Evaluation}
This section evaluates our ideas by comparing~\transmap~against state-of-the-art hardware prefetchers.
\subsection{Methodology}
\subsubsection{Simulator} We evaluate our model using the simulation framework released by ML Prefetching Competition based on ChampSim~\cite{ChampSim}. We train the model with memory traces of the last level cache(LLC) and test on LLC prefetcher by generating prefetching entries and insert in LLC. 
\subsubsection{Benchmarks} We use SPEC06 and SPEC17 to evaluate our model performance. For each application, we collect memory requests in LLC within 25 million instructions, using the first 20 million for training and the following 5 million for testing.

\subsubsection{Baseline}
We select state-of-the-art prefetchers: Best-Offset prefetcher~\cite{michaud2016best} and Irregular Stream Buffer (ISB) prefetcher~\cite{jain2013linearizing} as baselines to evaluate our model.
\subsubsection{Metrics} We evaluate our model by comparing their accuracy, coverage, MPKI (miss per kilo-instructions) improvement, and IPC (instructions per cycle) improvement.
\subsection{Simulation Results}
Figure~\ref{fig:res} shows the simulation results on accuracy, coverage, MPKI improvement, and IPC improvement comparing the our~\transmap~and baselines. Applications with index \textit{4xx} are from SPEC06 and those with index \textit{6xx} are from SPEC17.

~\transmap~achieves the highest average accuracy at 66.72\% while BO achieves 41.29\% and ISB achieves 61.79\%. Our model presents the highest coverage at 37.77\% that outperforms 30.58\% from BO and 22.26\% from ISB.~\transmap~presents the highest MPKI improvement, which means~\transmap~reduces 35.67\% of the misses per kilo-instructions, compared to 29.51\% and 22.51\% from BO and ISB respectively. Overall, ~\transmap~provides 20.55\% IPC improvement, which is higher than 19.07\% provided by BO prefetcher and 13.75\% provided by ISB prefetcher.

\section{Discussion}

\noindent\textbf{Prefetcher Preference} Different applications rely differently on temporal or spatial localities. While BO tracks the spatial correlation and ISB tracks the temporal correlation,~\transmap~learns from temporal sequences and predicts future accesses within a spatial range. This design makes~\transmap~ perform better for the average of all applications.

\noindent\textbf{Feasibility} The transformer model is more feasible for parallel computation in hardware implementation because there are no recurrent loops like in LSTM. Therefore, the Transformer inference latency is smaller comparing to LSTM in the same size. Besides, the bitmap labeling method can achieve long-horizon prefetching and can offset the latency of model inference.

\noindent\textbf{Online Retraining} While~\transmap~shows a powerful capacity in data prefetching, the performance relies on the size of the training and testing dataset. Specifically, without online updates, the prediction capacity will attenuate and both the accuracy and coverage will decay with an increasing number of testing instructions. To maintain the performance, online model updates are necessary and we will study this field in our future work.

\section{Conclusion}
In this paper, we have created a Transformer-based model for data prefetching. We use binary representation of absolute address as model input that solves the class explosion problem. We use the bitmap labeling method to collect unordered future block indexes for the current page, which solves the labeling problem and offsets computation latency for feasibility. The simulation results show that the proposed~\transmap~model achieves 35.67\% MPKI improvement and 20.55\% IPC improvement, higher than state-of-the-art BO prefetcher and ISB prefetcher.

\section*{Acknowledgements}
This work is supported by Air Force Research Laboratory grant number FA8750-18-S-7001, and National Science Foundation award number 1912680.


\bibliographystyle{IEEEtranS}
\bibliography{refs}

\end{document}